\documentclass[superscriptaddress,twocolumn,10pt,nofootinbib,aps,floatfix]{revtex4-1}
\usepackage{amssymb}
\usepackage{amsmath}
\usepackage{appendix}
\usepackage{times}
\usepackage{graphicx}
\usepackage{float}
\usepackage{subeqnarray}

\usepackage{color}
\usepackage{bm}

\begin{document}
\title{Hidden wallpaper fermion and third-order topological insulator}

\author{Ning Mao}
\author{Hao Wang}
\author{Ying Dai}
\email{daiy60@sdu.edu.cn}
\author{Baibiao Huang}
\author{Chengwang Niu}
\email{c.niu@sdu.edu.cn}
\affiliation
{School of Physics, State Key Laboratory of Crystal Materials, Shandong University, Jinan 250100, China}


\begin{abstract}
Nonsymmorphic symmetry can induce exotic wallpaper fermions, e.g., hourglass fermion, fourfold-degenerate Dirac fermion, and M{\"o}bius fermion, as commonly believed only in nonsymmorphic wallpaper groups. Here, we extend the notion of wallpaper fermions to symmorphic wallpaper groups, and remarkably uncover the emergence of long-awaited third-order topological insulators. The symmetry analysis and k $\cdot$ p models reveal that nonessential symmetries play an essential role for obtaining the previously overlooked hidden surface spectrum. Based on this, we present the hourglass fermion, fourfold-degenerate Dirac fermion, and M{\"o}bius fermion in the (001) surface of Tl$_4$XTe$_3$ (X = Pb/Sn) with a symmorphic wallpaper group $p4m$. Remarkably, 16 helical corner states reside on eight corners in Kramers pair, rendering the first real electronic material of third-order topological insulator. A time-reversal polarized octupole polarization is defined to uncover the nontrivial third-order topology, as is implemented by the 2$^{nd}$ and 3$^{rd}$ order Wilson loop calculations. Our results could considerably broaden the range of wallpaper fermions and lay the foundation for future experimental investigations of third-order topological insulators.
\end{abstract}

\maketitle
\date{\today}

\vspace{0.8cm}
\noindent{\bf \textcolor{red}{INTRODUCTION}}

\noindent
Much of the recent interest has been prompted by the symmetry-protected topological quantum states, and in particular, the  classification of topological electronic states is greatly enriched while including the crystalline symmetries~\cite{Chiu2016RMP,sptorder3,Po2017,Song2018,Watanabe2018,Tang2019SA}. A notable example is the topological crystalline insulator (TCI) protected by the symmorphic mirror or rotation symmetry, where twofold-degenerate surface fermions appear in particular surfaces preserving the corresponding symmorphic symmetry~\cite{fu2011prl,Hsieh,zhang2019topological,Hsu20192d}. Only recently, this topological classification starts to reach out to the nonsymmorphic symmetries, combinations of point group operations and fractional translations, such as the hourgalss fermion experimentally observed in KHg$X$ family ($X$ = As, Sb, Bi)~\cite{hourglass,khgsb}, and the M{\"o}bius insulators~\cite{mobiuskondo,Zhang136407}. Moreover, a fourfold-degenerate Dirac fermion emerges, in contrast to the twofold degeneracy of conventional TCIs and time-reversal symmetry ($\mathbb{T}$) protected topological insulators (TIs)~\cite{Kane,bhz,Zhang2009np}, on the surface of a topological insulating phase with multiple glide lines, representing an exception to fermion-doubling theorems~\cite{firstwallpaper}. Physically, the degeneracy and compatibility relation of surface states can be constrained to the irreducible co-representation of 17 two-dimensional (2D) wallpaper groups~\cite{firstwallpaper}, and the hourgalss fermion,  M{\"o}bius insulators, and fourfold-degenerate Dirac fermion has been well studied in the all four nonsymmorphic wallpaper groups, $pg$, $pmg$, $pgg$ and $p4g$~\cite{firstwallpaper,wallpaperjacs,mobiuskondo,Zhang136407,mobiustwist, offcentered,fangnonsymmor,nonsymmorphichigher,liucxnonsymmor}. However, similar symmetry-allowed band degeneracies in symmorphic wallpaper groups, which host also the multiple glide lines such as in $p4m$ and $cmm$, has not yet to be thoroughly investigated.

On the other hand, the crystalline symmetries open up a new horizon of topological bulk–boundary correspondence, extended with quantized quadrupole or octupole moments, revealing the presence of higher-order topological insulators (HOTIs) ~\cite{Benalcazar61,benalcazar2017electric,highorderfirst}. For which, the d-dimensional nth-order ($n\geq 2$) topological phase holds gapless state at (d-n)-dimensional boundary but gapped state otherwise. Therefore the unique topology of HOTIs is localized at hinges and/or corners with chiral or helical modes, which is distinguished by the nested Wilson loop, i.e., 2$^{nd}$ and 3$^{rd}$ order ones for second-order TIs (SOTIs) and third-order TIs (TOTIs), respectively~\cite{Benalcazar61,ni2020demonstration,highordergranpj}. Despite being actively explored, efforts have mainly focused on the SOTIs~\cite{higherordereuln2as2,Garcia2018,Imhof2018,Peterson2018,xue2019acoustic,ni2019observation, highorderbieuo,Yue2019,mnbiteprl1,mnbitenm1,nature4,Park216803,frankbi,kempkes2019robust, noguchi2021evidence,highordergranpj,highordergraprl,highordersplit}, which have been observed experimentally in both the metamaterials such as electrical circuit~\cite{Imhof2018}, microwave~\cite{Peterson2018}, and acoustics~\cite{xue2019acoustic,ni2019observation}, and the electronic materials such as Bi~\cite{frankbi} and Bi$_4$Br$_4$~\cite{noguchi2021evidence}. The TOTIs have only been limited to metamaterials, explicit electronic materials that enable the exploration of exotic quantum phenomena, including fractional corner charges~\cite{ highorderinvariants,highordersonginvariants} and filling anomaly~\cite{Peterson1114}, have so far been missing~\cite{Benalcazar61,Xue2019PRL,liu2020octupole,ni2020demonstration}.

\begin{figure*}
\centering
\includegraphics{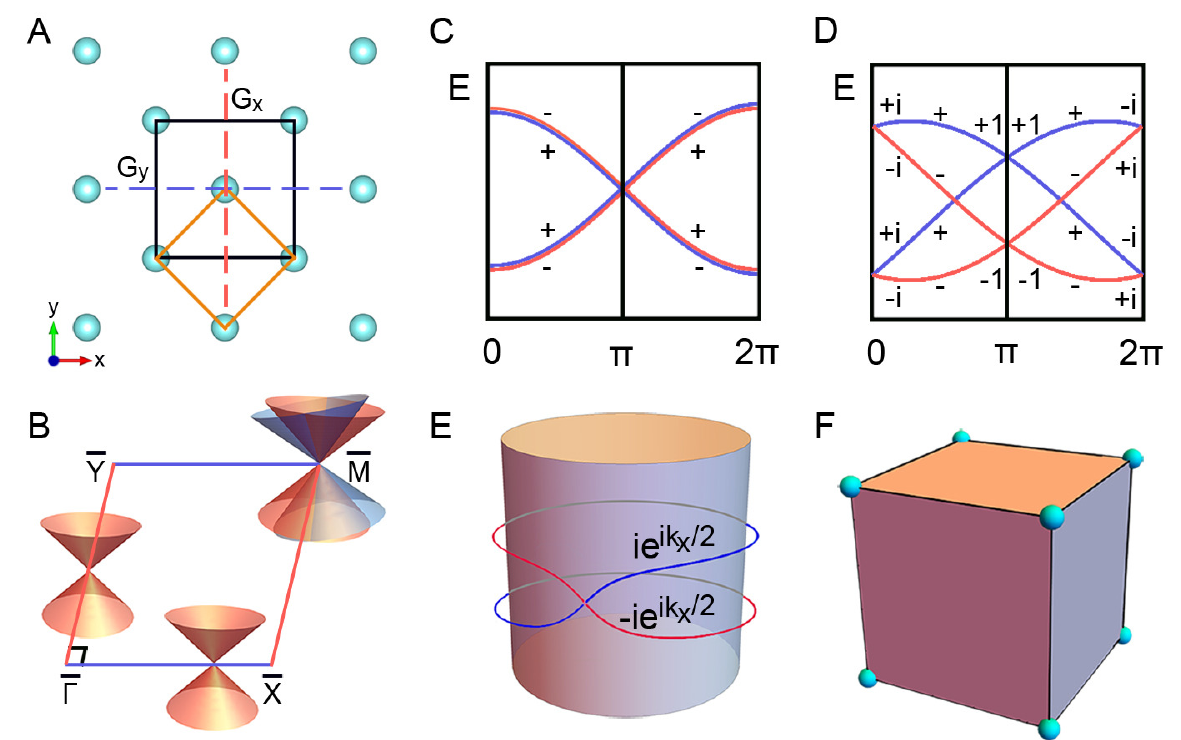}
\caption{
\textbf{Schematic demonstration for the hidden wallpaper fermions and third-order topology.} (\textbf{A}) Schematic of wallpaper group $p4m$ with two perpendicular glide lines $\mathbb{G}_{x/y}$ indicated with red/blue dashed lines. (\textbf{B}) Two categories of band crossings induced by $\mathbb{G}_{x/y}$ along the invariant paths, which are constrained by the commutation relations of $\mathbb{G}_{x/y}$ and $\mathbb{T}\mathbb{G}_{y/x}$. (\textbf{C}) When $\mathbb{G}_{x/y}$ anticommute with $\mathbb{T}\mathbb{G}_{y/x}$, a fourfold Dirac point appears at the zone boundary k = $\pi$, i.e., at M as in (\textbf{B}), (\textbf{D}) When $\mathbb{G}_{x/y}$ commute with $\mathbb{T}\mathbb{G}_{y/x}$, the band crossing is twofold degenerate and a hourglass-shaped dispersion with an internal partner switching for each quadruplet appears. (\textbf{E}) Due to the $\mathbb{G}_{x/y}$ eigenvalues $g_{\pm x/y} = \pm i e^{-ik_{y/x}/2}$, surface states can not go to the origin after a period of $2\pi$, but a periodicity of $4\pi$, featuring a M{\"o}bius fermion character. (\textbf{F}) Schematic of a helical third-order topological insulator with time-reversal bulk octupole moments.}
\label{Fig.1.}
\end{figure*}

Here, we demonstrate the realization of hidden wallpaper fermions and TOTIs in thallium compounds, Tl$_4X$Te$_3$ ($X$ = Pb/Sn), which have been overlooked as trivial insulators for years~\cite{tl4te3x4,nature1,nature2}. Correlated with two perpendicular glide mirrors in symmorphic wallpaper groups, hidden wallpaper fermions including hourgalss fermion, fourfold-degenerate Dirac fermion, and M{\"o}bius fermion emerge in the surface spectrum. However, unlike wallpaper fermions in nonsymmorphic wallpaper groups, the generalized $\mathbb{Z}_4$ invariant is trivial, which remarkably renders a gapped Wilson loop and allows us to implement the nested Wilson loop for the TOTIs. Indeed, the evaluated 3$^{rd}$ order Wilson loop for Tl$_4X$Te$_3$ are nonzero, confirming their nontrivial topology as the first family of TOTIs in electronic materials. Further analysis uncover that Tl$_4X$Te$_3$ are helical TOTIs with 16 nontrivial corner states reside on eight corners, very different from the reported chiral ones in metamaterials. Our results not only enrich the boundary of fundamental understanding of wallpaper fermion, but also provide a new route towards the realization of TOTIs. 

\vspace{0.8cm}
\noindent{\bf \textcolor{red}{RESULTS}}

\noindent{\bf \textcolor{black}{Wallpaper fermions in symmorphic wallpaper group}}

\noindent
It is well known that the band degeneracies are closely related with the symmetries in 17 2D space groups, i.e., wallpaper groups~\cite{firstwallpaper,liucxnonsymmor}. The wallpaper group $G$ can be thought of as a semi-direct product $G = T_{tp}\land Q$, where $T_{tp}$ is a pure translation group of Bravis lattice composed of infinite group elements and Q is a factor group used to distinguish the symmorphic and nonsymmorphic wallpaper groups. For a symmorphic wallpaper group, $Q$ is a pure point group composed of only the symmorphic symmetry operations, while the elements of $Q$ can be replaced by screw-axis rotations and/or glide lines for a nonsymmorphic wallpaper group. In fact, based on the definition of crystallographic wallpaper group, screw-axis rotations and/or glide lines are also possible to exist in symmorphic wallpaper groups. For example, six symmorphic wallpaper groups, $cm$, $cmm$, $p4m$, $p3m1$, $p31m$ and $p6m$, host the glide lines, and thus their surface states may have the similar manifolds as the nonsymmorphic wallpaper groups. Moreover, $cmm$ and $p4m$ can host two perpendicular glide lines, where fourfold-degenerate Dirac fermion is allowed to exist~\cite{lcxtnci,firstwallpaper}.

\begin{figure*}
\centering
\includegraphics{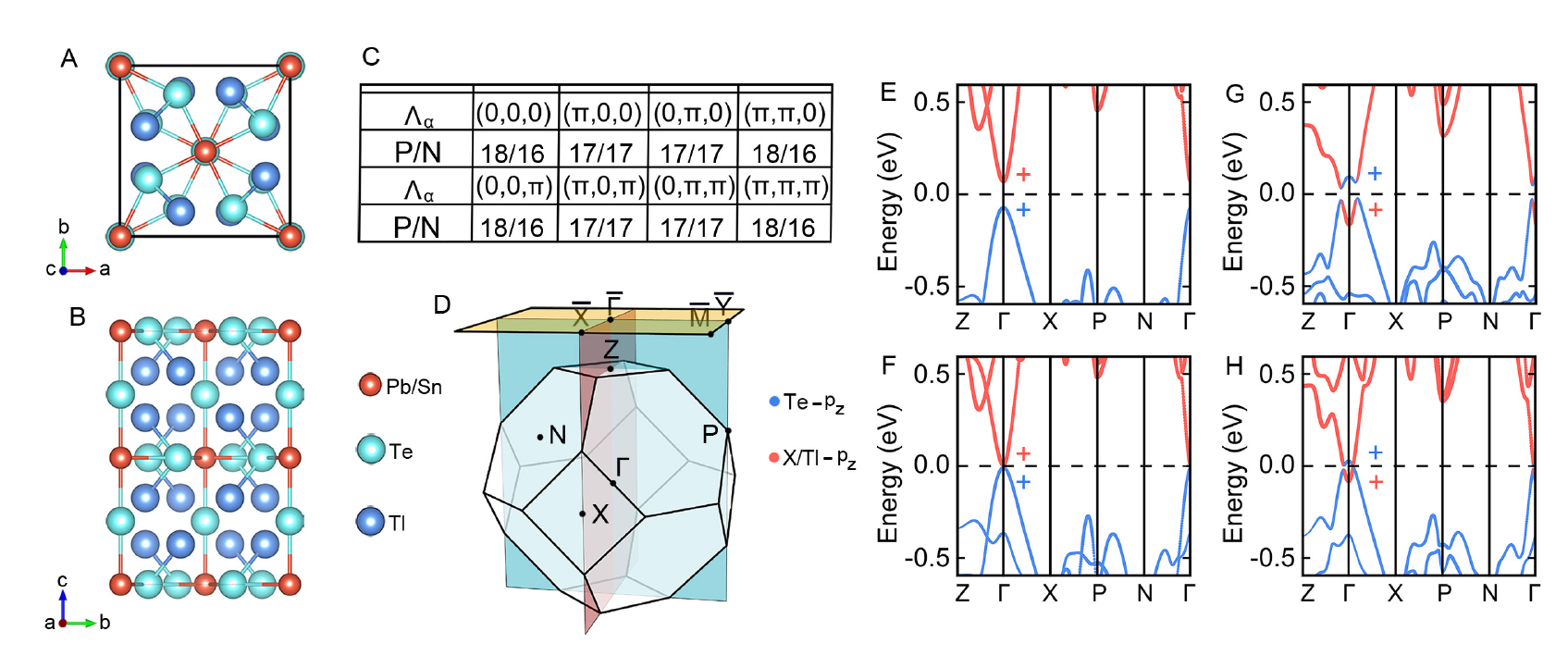}
\caption{
\textbf{Material candidates of hidden wallpaper fermion and third-order topological insulators.} (\textbf{A}) Top and (\textbf{B}) side views of Tl$_4X$Te$_3$ ($X$ = Pb/Sn) conventional unit cell. (\textbf{C}) Numbers of occupied bands with positive (P) and negative (N) parities at eight $\mathbb{T}$-invariant momenta ($\Lambda_{\alpha}$) with SOC. (\textbf{D}) Brillouin zone for primitive unit cell of Tl$_4X$Te$_3$, and projected 2D (001) Brillouin zone (yellow surface). Red and Blue planes are glide reflection invariant surfaces, which project into $\bar{\Gamma}\bar{X}$ and $\bar{\Gamma}\bar{Y}$ lines. Bulk band structures of (\textbf{E}), (\textbf{G}) Tl$_4$PbTe$_3$ and (\textbf{F}), (\textbf{H}) Tl$_4$SnTe$_3$ (\textbf{E}), (\textbf{F}), without and (\textbf{G}), (\textbf{H}) with SOC. The bands are orbitally weighted with the contribution of Te-p$_z$ and X/Tl-p$_z$ states. The Fermi level is indicated with a dashed line.}
\label{Fig.2.}
\end{figure*}

Taking the symmorphic wallpaper group $p4m$ as an example, we give a detailed description of the nonsymmorphic symmetry in symmorphic wallpaper group. $p4m$ is given by a semi-direct product of the translation group $T_{tp}$ and point group $4mm$, namely $p4m = T_{tp}  \land 4mm$. Indeed, there is no nonsymmorphic symmetry in point group $4mm$. Two symmorphic mirror lines, $\mathbb{M}_x$ ($x\rightarrow-x$) and $\mathbb{M}_y$ ($y\rightarrow-y$), are interesting to be mentioned. In addition, $T_{tp}$ consists of elements $\{E|n_1 \hat{t}_1+n_2\hat{t}_2+n_3(\hat{t}_1+\hat{t}_2)/2\}$, where $n_i ~(i=1,2,3)$ are integers, $\hat{t}_1, \hat{t}_2$ represent the lattice translation along x and y directions, respectively, and $(\hat{t}_1+\hat{t}_2)/2$ is the center translation that indicates a half-lattice translation when $n_3$ is odd. So, in general, nonsymmorphic symmetries are obtained by unique combination of $\mathbb{M}_{x/y}$ and $(\hat{t}_1+\hat{t}_2)/2$, which are glide lines $\mathbb{G}_{x/y} = \{\mathbb{M}_{x/y}|\frac{1}{2} \frac{1}{2}\}$, as illustrated in Fig.~\ref{Fig.1.}A. 

In crystal momentum space, the symmetry $\{\mathbb{D}|t\}$ can only protect degeneracies in the invariant lines/points of the Brillouin Zone that satisfy $\mathbb{D}k \rightarrow k$. In this invariant space, the Hamiltonian of system commutes with $\{\mathbb{D}|t\}$ and can be block diagonalized into two sectors with two opposite eigenvalues. For $\mathbb{G}_{x/y}$, the invariant lines are $\bar{\Gamma} \bar{Y}, \bar{X}\bar{M}/\bar{\Gamma} \bar{X}, \bar{Y}\bar{M}$, indicated with red/blue lines in Fig.~\ref{Fig.1.}B. Along the invariant lines, glide lines satisfy $\mathbb{G}_{x/y}^2 = - e^{-ik_{y/x}}$, so that one can get the eigenvalues as $g_{\pm x} = \pm i e^{-ik_y/2}$ and $g_{\pm y} = \pm i e^{-ik_x/2}$, which are $\pm 1$ at $\bar{X}$, $\bar{Y}$, $\bar{M}$ and $\pm i$ at $\bar{\Gamma}$. Besides, time-reversal symmetry ($\mathbb{T}$) imposes further constraints to the above four momenta, which satisfy $\mathbb{T}^2 = -1$, guaranteeing the Kramers degeneracy. The Kramers pairs at $\bar{X}$, $\bar{Y}$, and $\bar{M}$ have the same eigenvalues, while that at $\bar{\Gamma}$ have the opposite eigenvalues. Apart from unitary symmetry, it has to mention that $\bar{\Gamma} \bar{Y}, \bar{X}\bar{M}/\bar{\Gamma} \bar{X}, \bar{Y}\bar{M}$ are $\mathbb{T} \mathbb{G}_y/ \mathbb{T} \mathbb{G}_x$ invariant lines that may induce a higher degeneracy.

Along the above four glide invariant lines, the connectivities of bands are constrained by the commutation relations of $\mathbb{G}_{x/y}$ and $\mathbb{T}\mathbb{G}_{y/x}$, which can be classified into two categories. Figure~\ref{Fig.1.}C shows the first category with anticommutation relation $ \{\mathbb{G}_{x/y}, \mathbb{T}\mathbb{G}_{y/x}\} = 0$, the bands with opposite eigenvalues form a Kramers pair along the glide invariant lines. Moreover, at k = $\pi$, $\mathbb{G}_x$ and $\mathbb{G}_y$ anticommute with each other, i.e., $\{ \mathbb{G}_x, \mathbb{G}_y\} = 0$, and satisfy $\mathbb{G}_x^2 = \mathbb{G}_y^2 = +1$, a four-dimensional irreducible representations (Irreps) shows up that results in the fourfold-degenerate Dirac fermion at the $\bar{M}$ point. This can be further seen by examining the $k \cdot p$ Hamiltonian around the momenta $\bar{M}$,
\begin{align}
H_M = \tau_x (v_x \sigma_x k_x + v_y \sigma_y k_y).
\end{align}
At $\bar{M}$, the symmetry matrix can be written as $\mathbb{T} = i\sigma_yK$, $\mathbb{G}_x = \tau_y \sigma_x$, and $\mathbb{G}_y$ = $\tau_y \sigma_y$, where $\tau$ and $\sigma$ are Pauli matrices describing the sublattice and spin degrees of freedom. Clearly, the Dirac point remains intact for any glide lines allowed term, while for a $\mathbb{T}$-symmetric term $V_m$ = m $\tau_z$, it introduces a gap and turns the system into either a TI or a trivial insulator.

The second category shown in Fig.~\ref{Fig.1.}D exhibits a hourglass-shaped dispersion with the commutation relation $ [\mathbb{G}_{x/y}, \mathbb{T}\mathbb{G}_{y/x}] = 0$ along the momenta lines of $\bar{\Gamma} \bar{Y}$ and $\bar{\Gamma}\bar{X}$. The bands labeled with g$_{+ x/y}$ and g$_{- x/y}$ belong to different Irreps and will cross each other over the $\bar{\Gamma} \bar{Y}$ and $\bar{\Gamma} \bar{X}$, leading to a symmetry protected accidental degeneracy as shown in Fig.~\ref{Fig.1.}B. Due to any hybridization between different Irreps is not allowed by the glide line, $k\cdot p$ Hamiltonian around the crossing point can take the block diagonal forms as 
\begin{align}
H(\mathbf{k}) =  v_x k_x \sigma_x + v_y k_y \sigma_y,
\end{align}
where $\mathbf{k}$ represents the momenta of crossing point. Any $\mathbb{G}_{x/y}$ allowed terms can not break the degeneracy, but only shift the position of the degenerate point along the glide line, leading to a presence of $\mathbb{G}_{x/y}$-protected twofold-degenerate Dirac fermion. Remarkably, the phase of $\mathbb{G}_{y/x}$ entrust the birth of M{\"o}bius fermion as illustrated in Fig.~\ref{Fig.1.}E. That means, when one go round the $k_{x/y}$ direction about a period, $k_x$ $\rightarrow$ $k_x$ + 2$\pi$, the eigenstates switch places, and we need go around 4$\pi$ to return the origin.

Having demonstrated the possibility of the similar symmetry-allowed band degeneracies in $p4m$ (called as hidden wallpaper fermion) to that of the wallpaper fermion in nonsymmorphic $pgg$ and $p4g$, we aim now at showing that their topology can be totally different, and remarkably, a third-order TI is obtained. In the presence of $\mathbb{T}$, zero-dimensional corner states originate from the time-reversal polarized octupole moment, which are quantized to 0 or 1/2 by the crystallographic symmetries. As presented in Supplementary Material, together with inversion symmetry ($\mathbb{I}$),  mirror and/or glide symmetries along three cartesian directions have to participate in. Among the 4 wallpaper groups that host fourfold degenerate Dirac fermion, only $p4m$ and $p4g$ reside in all of the necessary symmetries to get the quantized time-reversal polarized octupole moment. However, the fact that existence of hinge states in $p4g$ as reported before makes the third-order topology can only survive in the $p4m$.

\begin{figure*}
\centering
\includegraphics{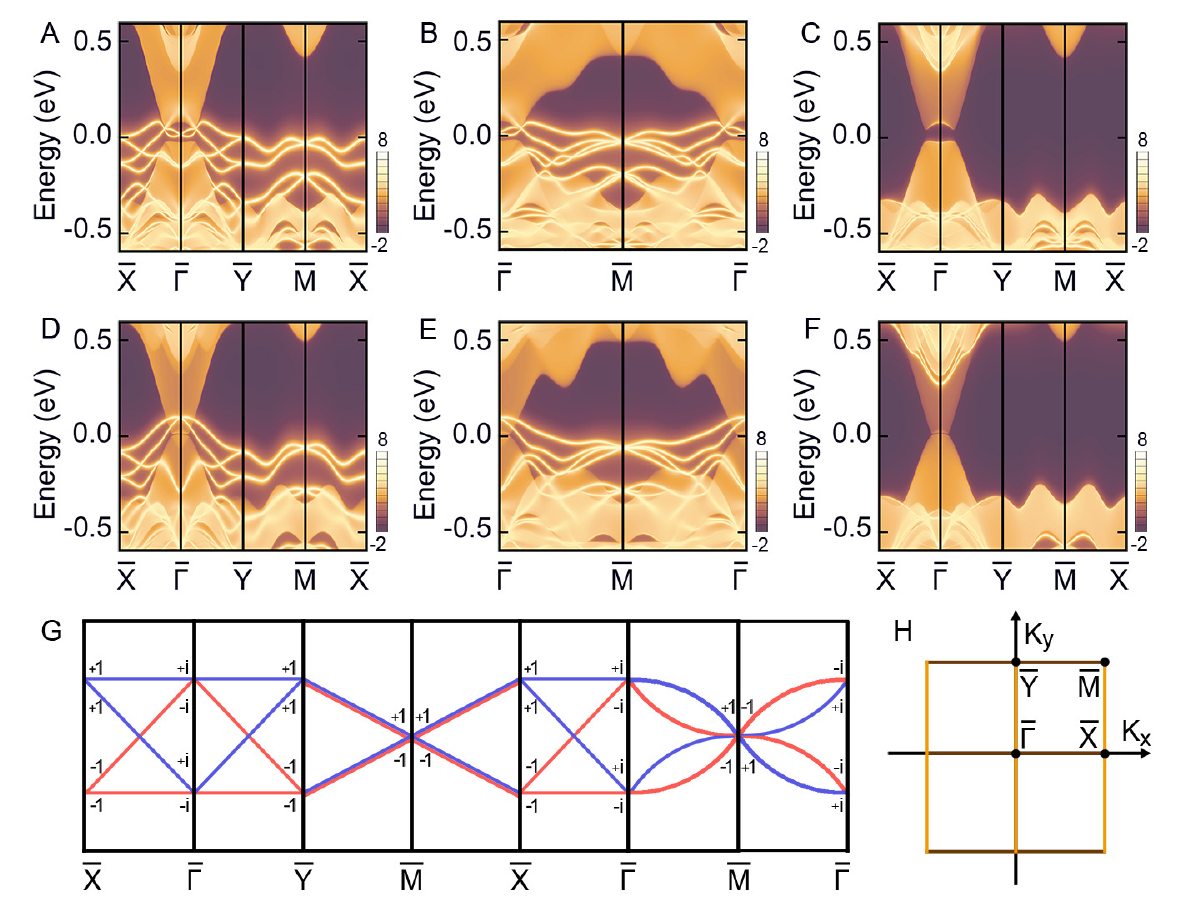}
\caption{
\textbf{Surface spectrum for hidden wallpaper fermion.} Band structures for (001) surface of (\textbf{A})-(\textbf{C}) Tl$_4$PbTe$_3$ and (D)-(F), Tl$_4$SnTe$_3$ calculated using surface Green’s functions. The Fermi level is set to zero. (\textbf{A}), (\textbf{B}), (\textbf{D}) and (\textbf{E}) The top surface displays hourglass fermions along the glide-symmetric lines $\bar{\Gamma}\bar{X}/\bar{\Gamma}\bar{Y}$ and the fourfold Dirac fermion at the $\bar{M}$ point, while (\textbf{C}), (\textbf{F}) there is no surface bands on the bottom surface, suggesting the M{\"o}bius fermion between opposite surfaces. (\textbf{G}) Schematic of hidden wallpaper fermions along the glide-invariant lines $\mathbb{G}_{x/y}$. The labels indicate the corresponding eigenvalues $\pm i e^{-ik_{x/y}/2}$. (\textbf{H}) Brillouin zone of the (001)-surface with $\mathbb{G}_x$/$\mathbb{G}_y$ invariant lines are indicated by yellow/brown lines.}
\label{Fig.3.}
\end{figure*}

\vspace{0.8cm}
\noindent{\bf \textcolor{black}{Material candidates}}

\noindent
For the materials realization, we focus on the ternary thallium compounds Tl$_4X$Te$_3$($X$ = Pb/Sn), which are known to be advanced thermoelectric and optoelectronic materials~\cite{tl4te3x1,tl4te3x3}. They crystallize in the tetragonal structure with a space group of $I4/mcm$ as shown in Figs.~\ref{Fig.2.}A and~\ref{Fig.2.}B~\cite{tl4te3x2}. There are 32 atoms in a conventional cell, with 16 Tl atoms located at Wyckoff position $16I$, 4 $X$ atoms located at Wyckoff position $4c$, and 12 Te atoms located at two Wyckoff positions $4a$ and $8h$~\cite{wyckoff}, respectively. The optimized lattice constants are $a = 8.98/8.93$ \AA~ and $c = 13.43/13.35$ \AA~ for Tl$_4X$Te$_3$ with $X$ = Pb/Sn, almost the same as with the experimental values~\cite{tl4te3x1}. Their wallpaper groups of (001)-projection, (100)-projection, and (110)-projection are $p4m$, $pmm$, and $pmm$, respectively, with both $pmm$ and $p4m$ are symmorphic wallpaper groups. The $pmm$ contains only the symmorphic mirror and rotation symmetries, while, interestingly, the $p4m$ manifests the existence of glide lines, which may give birth to hidden wallpaper fermion and higher-order topology as discussed above. 

The orbitally resolved band structures of Tl$_4X$Te$_3$ without and with SOC are illustrated in Figs.~\ref{Fig.2.}E -~\ref{Fig.2.}H. In the absence of SOC, Tl-$p_z$ and Pb/Sn-$p_z$ orbitals contribute to the conduction band minimum (CBM), while the valence band maximum (VBM) is dominated by Te-$p_z$
orbitals with direct energy gaps of 142 meV and 16 meV for Tl$_4$PbTe$_3$ and Tl$_4$SnTe$_3$, respectively, as illustrated in Figs.~\ref{Fig.2.}E and ~\ref{Fig.2.}F. Switching on SOC, the insulating character is preserved for both Tl$_4$PbTe$_3$ and Tl$_4$SnTe$_3$ with corresponding energy gaps of 83 meV and 20 meV, which are similar to previous predictions~\cite{nature1,nature2}, and remarkably, an inversion of the orbital characters around the  $\Gamma$ point occurs as shown in Figs.~\ref{Fig.2.}G and ~\ref{Fig.2.}H. However, qualitatively different from the reported $s-p$ or $p-p$ band inversions in TIs and TCIs, the parity of Bloch state forms the CBM and VBM at $\Gamma$ are all positive, i.e., there is no parity exchange between occupied and unoccupied bands in the SOC-induced $p-p$ band inversion for Tl$_4X$Te$_3$.

\begin{figure*}
\centering
\includegraphics{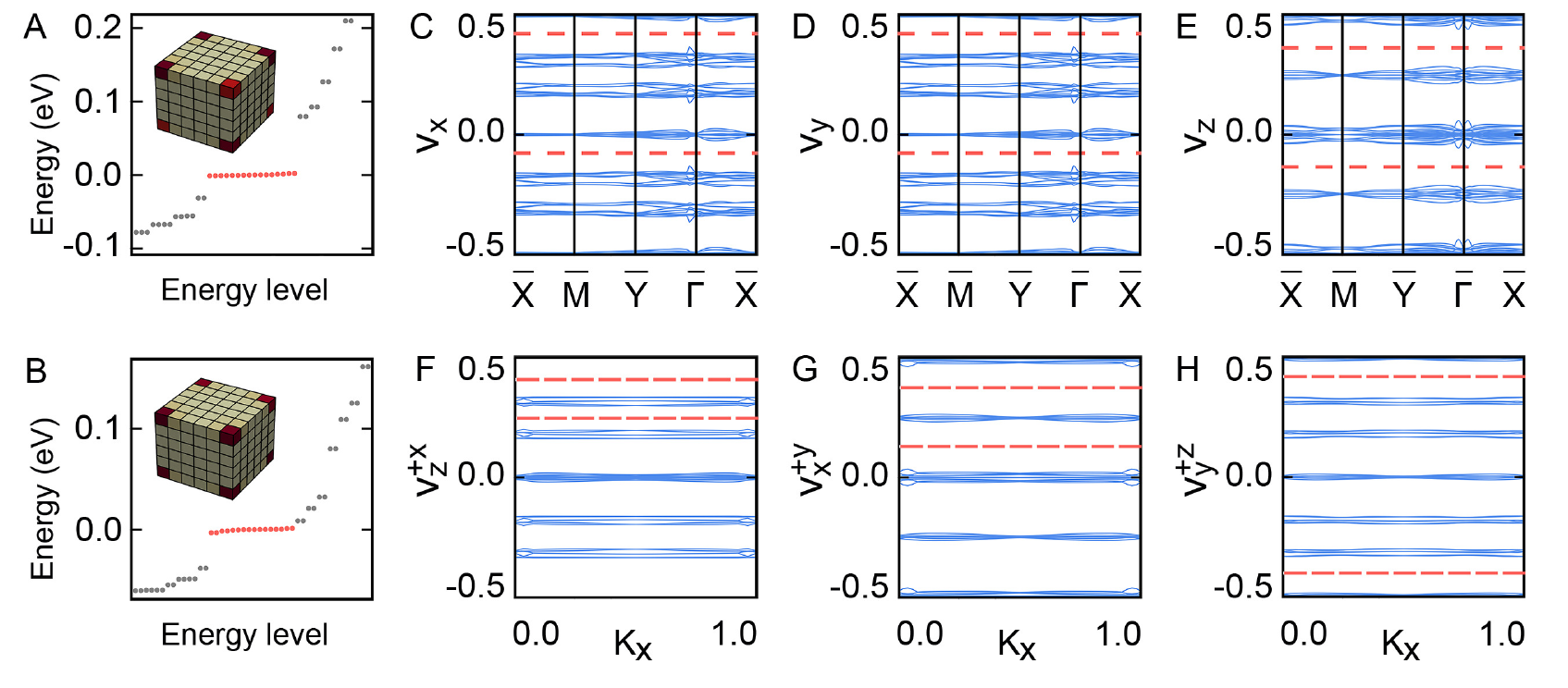}
\caption{
\textbf{Nontrivial helical corner states.} Energy levels of a finite lattice composed of 6 $\times$ 6 $\times$ 6 conventional unit cells for (\textbf{A}) Tl$_4$PbTe$_3$ and (\textbf{B}) Tl$_4$SnTe$_3$. There are 16 degenerate corner states as indicated with red color. Insets show the probability of one corner state, where we sum all contributions of probability in one unit cell and each unit cell is represented with a cubic. (\textbf{C})-(\textbf{E}) 1$^{st}$ order wilson bands $v_x$, $v_y$, and $v_z$ of the occupied bulk bands. The generalized "Fermi levels" are indicated with red dash lines, where we take the 1$^{st}$ order wilson bands among the two red lines as occupied 1$^{st}$ order wilson bands. (\textbf{F})-(\textbf{H}) 2$^{nd}$ order wilson bands $v_z^{+x}$, $v_x^{+y}$, and $v_y^{+z}$ of the selective occupied 1$^{st}$ order wilson bands. The generalized "Fermi levels" are indicated with red dash lines, where we take the 2$^{nd}$ order wilson bands among the two red lines as occupied 2$^{nd}$ order wilson bands.}
\label{Fig.4.}
\end{figure*}

To ensure the parity judgement, we implement the direct calculations of topological $\mathbb{Z}_2$ invariants $\nu_0;(\nu_1, \nu_2, \nu_3)$ according to the parity of each Kramers pair at eight $\mathbb{T}$-invariant momenta ($\Lambda_{\alpha}$)~\cite{z2invariant}. The number of odd and even bands are listed in Fig.~\ref{Fig.2.}C, and therefore the parity products at eight $\Lambda_{\alpha}$ are given as (+, -, -, +, +, -, -, +), that results in $\nu_0$ = 0 and ($\nu_1, \nu_2, \nu_3$) = (0, 0, 0), revealing the trivial $\mathbb{Z}_2$ property. In addition, as reported for TCIs with wallpaper fermions, one can always use the Wilson loop eigenvalues along the glide symmetry preserved surface to define two generalized $\mathbb{Z}_4$ invariants, $(\chi_x, \chi_y)$, which can be evaluated by the crossing of each glide sector and an arbitrary horizontal line~\cite{firstwallpaper}. As presented in  Figs.~\ref{Fig.4.}C-~\ref{Fig.4.}E, trivial generalized $\mathbb{Z}_4$ indices with ($\chi_x, \chi_y$) = (0, 0) are obtained, revealing the different topology from wallpaper fermion in $p4g$ that is one of major characters of our defined hidden wallpaper fermion.

Actually, the calculated $\mathbb{Z}_2$ and $\mathbb{Z}_4$ invariants agree with the previous works that classified Tl$_4X$Te$_3$ into trivial insulators~\cite{nature1,nature2,tl4te3x4}, here remarkably, we further detail their topology, and identify them as the first realistic electronic material realizations of TOTIs with exotic hidden wallpaper fermions and corner states. To prove the nontrivial topology, similar to the concept of time-reversal polarization\cite{z2bump}, we define a time-reversasl polarized octupole polarization based on the Wilson loop and nested Wilson loop (see Supplementary Sections 1 for more details). Firstly, $1^{st}$ order Wilson loop along x, y and z-direction are calculated with 72 occupied bulk bands as shown in Figs.~\ref{Fig.4.}C - ~\ref{Fig.4.}E. We chose half of the 1$^{st}$ Wilson bands to do the $2^{nd}$ order Wilson loop by specifying two generalized "Fermi levels", indicated with red dashed lines. The spectrum of $2^{nd}$ order Wilson bands are shown in Figs.~\ref{Fig.4.}F - ~\ref{Fig.4.}H. Since the $2^{nd}$ order Wilson bands are still gapped, we can implement the calculations of $3^{rd}$ order Wilson loop under six occupied $2^{nd}$ order Wilson bands. All of the six $3^{rd}$ order Wilson bands show us a nontrivial value of 0.5, leading to a vanishing octupole polarization. However, for the calculation of time-reversal polarized octupole polarization, we only have to take one eigenvalue of each $3^{rd}$ order Wilson Kramers pair. The result show us
\begin{align}
(\tilde{p}_{x,+}^{+z,+y}, \tilde{p}_{y,+}^{+x,+z}, \tilde{p}_{z,+}^{+y,+x}) = (\frac{1}{2}, \frac{1}{2}, \frac{1}{2}),
\end{align}
demonstrating the TOTI nature of Tl$_4X$Te$_3$.

\vspace{0.8cm}
\noindent{\bf \textcolor{black}{Nontrivial hidden wallpaper fermions}}

\noindent
The non-trivial (d-1) dimensional topology of hidden wallpaper fermions of Tl$_4X$Te$_3$ originate from two perpendicular glide planes $g_x = \{M_x|\frac{1}{2} \frac{1}{2} 0 \}$ and $g_y = \{M_y|\frac{1}{2} \frac{1}{2} 0\}.$ Those glide planes can only be on effects in their symmetry-preserving crystal surface, which need to perpendicular to the mirror planes and also invariant under the translation along the glide translation direction. The only surface that satisfies those conditions is the (001) surface with wallpaper group $p4m$. Figures~\ref{Fig.3.}A and~\ref{Fig.3.}B (\ref{Fig.3.}D and \ref{Fig.3.}E) display the nontrivial surface states with hourgalss fermion and fourfold-degenerate Dirac fermion for Tl$_4$PbTe$_3$ (Tl$_4$SnTe$_3$), and Fig.~\ref{Fig.3.}C (\ref{Fig.3.}F) presents the opposite (001) surface states to clearly show the M{\"o}bius fermion. To understand the nature of these band crossings, we have to check the glide eigenvalues of the surface bands as shown in Fig.~\ref{Fig.3.}G.

\ (1) At the $\bar{X}$, $\bar{Y}$, and $\bar{M}$ points, glide eigenvalues are always $\pm$1 and thus all bands are doubly-degenerate with the same glide eigenvalues due to the limitation of $\mathbb{T}$. On the contrary, the glide eigenvalues are $\pm$i at $\bar{\Gamma}$, leading to a double-degeneracy with different glide eigenvalues. 

\ (2) Along $g_x$ ($g_y$) invariant line $\bar{\Gamma}\bar{Y}$ ($\bar{\Gamma}\bar{X}$), one of the Kramers pair at $\bar{\Gamma}$ changes its partner with one of the Kramer pairs at $\bar{Y}$ ($\bar{X}$), constituting so-called hourglass surface band structure. During the process of partner switching, a twofold-degenerate Dirac point emerge for the glide invariant line, which is not pinned to any momenta.

\ (3) Along $g_x$ ($g_y$) invariant line $\bar{M} \bar{X}$ ($\bar{M} \bar{Y}$), bands are always doubly-degenerate instead of forming hourglass structure, which is caused by commutation relation of $ g_x g_y = -g_y g_x  t_x t_{-y}$. That means, a state $| \psi_{\mathbf{k}} \rangle$ with $g_{+x}$ eigenvalues will be degenerate with a state $\mathbb{T} g_y| \psi_{\mathbf{k}} \rangle$ with opposite $g_{-x}$ eigenvalues. Thus all sets of degenerate pairs will have the zero glide eigenvalues, leading to only one kind of Irreps. It has to emphasize that there is no chance for formation of fourfold-degenerate bands along such lines due to the avoided crossing of the same Irreps.

\ (4) A fourfold-degenerate Dirac point can only be obtained at the $\bar{M}$ point due to the anticommutation relation of $\{g_x, g_y\} = 0$. This means that, a state $| \psi_{\mathbf{k}} \rangle$, eigenstate of $g_x$, will be degenerate with three eigenstates of $\mathbb{T} | \psi_{\mathbf{k}} \rangle$, $g_y| \psi_{\mathbf{k}} \rangle$, and $\mathbb{T} g_y| \psi_{\mathbf{k}} \rangle$.

\ (5) The fourfold-degenerate Dirac point splits into quadruplets away from $\bar{M}$ along the $\bar{M}\bar{\Gamma}$ because of either $g_x$ or $g_y$ is broken. Bands can not be degenerate, but only split into single branches. 

\ (6) In contrary to the nontrivial surface states of top surface, the bottom surface shows us nothing, which is also called as a M{\"o}bius twist. Such twsit is originated from the intrinsic translation part of glide planes. As we can see from the glide eigenvalues of $ g_{\pm x} = \pm i e^{-ik_y/2}$ and $g_{\pm y} = \pm i e^{-ik_x/2}$, one band with eigenvalue $g_{+ x/y}$ will turn into $g_{- x/y}$ instead of turning into itself like a mirror plane. Such unusual connectivity give rise to the M{\"o}bius twist character, which is totally distinct from the symmorphic symmetry protected TCIs.

\vspace{0.8cm}
\noindent{\bf \textcolor{black}{Nontrivial helical corner states}}

\noindent
At last, we investigate the emergence of nontrivial corner states in our systems, which are the hallmarks of 3D TOTIs, by using the maximally localized Wannier functions that can reproduce the band dispersions of Tl$_4$XTe$_3$ very accurately. Firstly, we construct a finite lattice composed of 6 $\times$ 6 $\times$ 6 conventional unit cells as exemplified in Supplementary Materials. Notably, as shown in Figs.~\ref{Fig.4.}A and ~\ref{Fig.4.}B, 16 nearly degenerate states arise around the Fermi level, highlighted in red color. Inset presents the corresponding real-space distribution of one corner state. Indeed, the wave functions of this state are localized almost at the 8 corners and vanish in other regions, which is a significant signal for the corner states. It is interesting to emphasize that 16 corner states are helical, different from the previous quantized octupole TIs with chiral modes~\cite{Benalcazar61,Xue2019PRL,liu2020octupole,ni2020demonstration}. Due to the degeneracy of 16 in-gap states, the half-filling condition cannot be satisfied as long as $\mathbb{P}$ is preserved, which is known as the filling anomaly. To resolve filling anomaly, extra electrons or holes need to fullfill the valence band completely, which result in accumulations of electrons or holes.

\vspace{0.8cm}
\noindent{\bf \textcolor{red}{DISCUSSION}}

\noindent
In summary, We have proposed Tl$_4$XTe$_3$ (X = Pb/Te) as promising material candidates for both hidden wallpaper fermions and TOTIs using k$\cdot$p models and ab-initio calculations. Although in a symmorphic wallpaper group, the exotic hourglass fermion, four-fold degenerate Dirac fermion, and M{\"o}bius fermion can still emerge on the (001) surface. Different from the wallpaper fermions predicted from a nonsymmorphic wallpaper group, our proposed materials have a trivial $Z_4$ invariant that is in agreement with the  results of topological quantum chemistry. Such a trivial $Z_4$ invariant bring us a gapped wilson loop, making the Tl$_4$XTe$_3$ to be TOTIs. Due to the existence of $\mathbb{T}$, corner states come in Kramers pair and have a zero octupole polarization, which is different from the quantized otupole TIs achieved in metamaterials~\cite{liu2020octupole,ni2020demonstration}. Each one of Kramers pair has a nontrivial otupole polarization, leading to a nontrivial time-reversal polarized octupole polarization. It is interesting to emphasize that our proposal for hidden wallpaper fermion broden the survey of wallpaper fermion. Due to the existence of center vector, symmorphic wallpaper groups like $cm$, $cmm$, $p4m$, $p3m1$, $p31m$, and $p6m$ can possess glide mirror planes, which make them possible to show wallpaper fermions. Such hidden wallpaper fermions may exhibit exotic nontrivial characters like the third-order topology. Our findings may motivate the formulation of novel strategies in finding topological phenomena and exotic topological states for spintronics applications.

\vspace{0.8cm}
\noindent{\bf \textcolor{red}{MATERIALS AND METHODS}}

\noindent{\bf \textcolor{black}{First-principles calculations}}

\noindent
The first-principles calculations are carried out in the framework of generalized gradient approximation (GGA)  with Perdew–Burke–Ernzerhof (PBE)~\cite{pbevasp} functionals using the Vienna Ab initio simulation package (VASP)~\cite{vasp} and the full-potential linearized augmented-plane-wave method using the FLEUR code~\cite{fleur}. The self-consistent total energy was evaluated with a 12$\times$12$\times$8 k-point mesh, and the cutoff energy for the plane-wave basis set was 500 eV. The maximally localized Wannier functions (MLWFs) are constructed using the Wannier90 code~\cite{wannier90} in conjunction with the FLEUR packag~\cite{fluer1,fluer2} and the edge states calculations are using WannierTools~\cite{wanniertools}, where we choose p orbital of Pb/Sn, Te and Tl atoms to construct MLWFs.

\vspace{0.8cm}
\def\bibsection{\noindent{\bf \textcolor{red}{REFERENCES AND NOTES}}}


\begin{thebibliography}{10}
\expandafter\ifx\csname url\endcsname\relax
  \def\url#1{\texttt{#1}}\fi
\expandafter\ifx\csname urlprefix\endcsname\relax\def\urlprefix{URL }\fi
\providecommand{\bibinfo}[2]{#2}
\providecommand{\eprint}[2][]{\url{#2}}

\bibitem{Chiu2016RMP}
\bibinfo{author}{C.-K. Chiu}, \bibinfo{author}{J.~C.~Y. Teo},
  \bibinfo{author}{A.~P. Schnyder}, \bibinfo{author}{S.~Ryu}.
\newblock \bibinfo{title}{Classification of topological quantum matter with
  symmetries}.
\newblock \emph{\bibinfo{journal}{Rev. Mod. Phys.}}
  \textbf{\bibinfo{volume}{88}}, \bibinfo{pages}{035005}
  (\bibinfo{year}{2016}).

\bibitem{sptorder3}
\bibinfo{author}{C.~Wang}, \bibinfo{author}{C.-H. Lin},
  \bibinfo{author}{M.~Levin}.
\newblock \bibinfo{title}{Bulk-boundary correspondence for three-dimensional
  symmetry-protected topological phases}.
\newblock \emph{\bibinfo{journal}{Phys. Rev. X}} \textbf{\bibinfo{volume}{6}},
  \bibinfo{pages}{021015} (\bibinfo{year}{2016}).

\bibitem{Po2017}
\bibinfo{author}{H.~C. Po}, \bibinfo{author}{A.~Vishwanath},
  \bibinfo{author}{H.~Watanabe}.
\newblock \bibinfo{title}{{Symmetry-based Indicators of Band Topology in the
  230 Space Groups}}.
\newblock \emph{\bibinfo{journal}{Nat. Commun.}} \textbf{\bibinfo{volume}{8}},
  \bibinfo{pages}{50} (\bibinfo{year}{2017}).

\bibitem{Song2018}
\bibinfo{author}{Z.~Song}, \bibinfo{author}{T.~Zhang},
  \bibinfo{author}{Z.~Fang}, \bibinfo{author}{C.~Fang}.
\newblock \bibinfo{title}{{Quantitative mappings between symmetry and topology
  in solids}}.
\newblock \emph{\bibinfo{journal}{Nat. Commun.}} \textbf{\bibinfo{volume}{9}},
  \bibinfo{pages}{3530} (\bibinfo{year}{2018}).

\bibitem{Watanabe2018}
\bibinfo{author}{H.~Watanabe}, \bibinfo{author}{H.~C. Po},
  \bibinfo{author}{A.~Vishwanath}.
\newblock \bibinfo{title}{{Structure and topology of band structures in the
  1651 magnetic space groups}}.
\newblock \emph{\bibinfo{journal}{Sci. Adv.}} \textbf{\bibinfo{volume}{4}},
  \bibinfo{pages}{eaat8685} (\bibinfo{year}{2018}).

\bibitem{Tang2019SA}
\bibinfo{author}{F.~Tang}, \bibinfo{author}{H.~C. Po},
  \bibinfo{author}{A.~Vishwanath}, \bibinfo{author}{X.~Wan}.
\newblock \bibinfo{title}{{Topological materials discovery by large-order
  symmetry indicators}}.
\newblock \emph{\bibinfo{journal}{Sci. Adv.}} \textbf{\bibinfo{volume}{5}},
  \bibinfo{pages}{eaau8725} (\bibinfo{year}{2019}).

\bibitem{fu2011prl}
\bibinfo{author}{L.~Fu}.
\newblock \bibinfo{title}{{Topological crystalline insulators}}.
\newblock \emph{\bibinfo{journal}{Phys. Rev. Lett.}}
  \textbf{\bibinfo{volume}{106}}, \bibinfo{pages}{106802}
  (\bibinfo{year}{2011}).

\bibitem{Hsieh}
\bibinfo{author}{T.~H. Hsieh}, \bibinfo{author}{H.~Lin},
  \bibinfo{author}{J.~Liu}, \bibinfo{author}{W.~Duan},
  \bibinfo{author}{A.~Bansil}, \bibinfo{author}{L.~Fu}.
\newblock \bibinfo{title}{{Topological crystalline insulators in the SnTe
  material class}}.
\newblock \emph{\bibinfo{journal}{Nat. Commun.}} \textbf{\bibinfo{volume}{3}},
  \bibinfo{pages}{982} (\bibinfo{year}{2012}).

\bibitem{zhang2019topological}
\bibinfo{author}{T.~Zhang}, \bibinfo{author}{C.~Yue},
  \bibinfo{author}{T.~Zhang}, \bibinfo{author}{S.~Nie},
  \bibinfo{author}{Z.~Wang}, \bibinfo{author}{C.~Fang},
  \bibinfo{author}{H.~Weng}, \bibinfo{author}{Z.~Fang}.
\newblock \bibinfo{title}{Topological crystalline insulators with {$C_2$}
  rotation anomaly}.
\newblock \emph{\bibinfo{journal}{Phys. Rev. Res.}}
  \textbf{\bibinfo{volume}{1}}, \bibinfo{pages}{012001} (\bibinfo{year}{2019}).

\bibitem{Hsu20192d}
\bibinfo{author}{C.-H. Hsu}, \bibinfo{author}{X.~Zhou},
  \bibinfo{author}{Q.~Ma}, \bibinfo{author}{N.~Gedik},
  \bibinfo{author}{A.~Bansil}, \bibinfo{author}{V.~M. Pereira},
  \bibinfo{author}{H.~Lin}, \bibinfo{author}{L.~Fu}, \bibinfo{author}{S.-Y.
  Xu}, \bibinfo{author}{T.-R. Chang}.
\newblock \bibinfo{title}{{ Purely rotational symmetry-protected topological
  crystalline insulator $\alpha$-Bi$_4$Br$_4$ }}.
\newblock \emph{\bibinfo{journal}{2D Mater.}} \textbf{\bibinfo{volume}{6}},
  \bibinfo{pages}{031004} (\bibinfo{year}{2019}).

\bibitem{hourglass}
\bibinfo{author}{Z.~Wang}, \bibinfo{author}{A.~Alexandradinata},
  \bibinfo{author}{R.~J. Cava}, \bibinfo{author}{B.~A. Bernevig}.
\newblock \bibinfo{title}{Hourglass fermions}.
\newblock \emph{\bibinfo{journal}{Nature}} \textbf{\bibinfo{volume}{532}},
  \bibinfo{pages}{189} (\bibinfo{year}{2016}).

\bibitem{khgsb}
\bibinfo{author}{J.~Ma}, \bibinfo{author}{C.~Yi}, \bibinfo{author}{B.~Lv},
  \bibinfo{author}{Z.~Wang}, \bibinfo{author}{S.~Nie},
  \bibinfo{author}{L.~Wang}, \bibinfo{author}{L.~Kong},
  \bibinfo{author}{Y.~Huang}, \bibinfo{author}{P.~Richard},
  \bibinfo{author}{P.~Zhang}, et~al.
\newblock \bibinfo{title}{Experimental evidence of hourglass fermion in the
  candidate nonsymmorphic topological insulator in {KHgSb}}.
\newblock \emph{\bibinfo{journal}{Sci. Adv.}} \textbf{\bibinfo{volume}{3}},
  \bibinfo{pages}{e1602415} (\bibinfo{year}{2017}).

\bibitem{mobiuskondo}
\bibinfo{author}{P.-Y. Chang}, \bibinfo{author}{O.~Erten},
  \bibinfo{author}{P.~Coleman}.
\newblock \bibinfo{title}{M{\"o}bius kondo insulators}.
\newblock \emph{\bibinfo{journal}{Nat. Phys.}} \textbf{\bibinfo{volume}{13}},
  \bibinfo{pages}{794--798} (\bibinfo{year}{2017}).

\bibitem{Zhang136407}
\bibinfo{author}{R.-X. Zhang}, \bibinfo{author}{F.~Wu},
  \bibinfo{author}{S.~Das~Sarma}.
\newblock \bibinfo{title}{M\"obius insulator and higher-order topology in
  {MnBi$_{2n}$Te$_{3n+1}$}}.
\newblock \emph{\bibinfo{journal}{Phys. Rev. Lett.}}
  \textbf{\bibinfo{volume}{124}}, \bibinfo{pages}{136407}
  (\bibinfo{year}{2020}).

\bibitem{Kane}
\bibinfo{author}{C.~L. Kane}, \bibinfo{author}{E.~J. Mele}.
\newblock \bibinfo{title}{Quantum spin {Hall} effect in graphene}.
\newblock \emph{\bibinfo{journal}{Phys. Rev. Lett.}}
  \textbf{\bibinfo{volume}{95}}, \bibinfo{pages}{226801}
  (\bibinfo{year}{2004}).

\bibitem{bhz}
\bibinfo{author}{B.~A. Bernevig}, \bibinfo{author}{T.~L. Hughes},
  \bibinfo{author}{S.-C. Zhang}.
\newblock \bibinfo{title}{{Quantum spin Hall effect and topological phase
  transition in HgTe quantum wells}}.
\newblock \emph{\bibinfo{journal}{Science}} \textbf{\bibinfo{volume}{314}},
  \bibinfo{pages}{1757--1761} (\bibinfo{year}{2006}).

\bibitem{Zhang2009np}
\bibinfo{author}{H.~Zhang}, \bibinfo{author}{C.-X. Liu}, \bibinfo{author}{X.-L.
  Qi}, \bibinfo{author}{X.~Dai}, \bibinfo{author}{Z.~Fang},
  \bibinfo{author}{S.-C. Zhang}.
\newblock \bibinfo{title}{{Topological insulators in Bi$_2$Se$_3$, Bi$_2$Te$_3$
  and Sb$_2$Te$_3$ with a single Dirac cone on the surface}}.
\newblock \emph{\bibinfo{journal}{Nat. Phys.}} \textbf{\bibinfo{volume}{5}},
  \bibinfo{pages}{438--442} (\bibinfo{year}{2009}).

\bibitem{firstwallpaper}
\bibinfo{author}{B.~J. Wieder}, \bibinfo{author}{B.~Bradlyn},
  \bibinfo{author}{Z.~Wang}, \bibinfo{author}{J.~Cano},
  \bibinfo{author}{Y.~Kim}, \bibinfo{author}{H.-S.~D. Kim},
  \bibinfo{author}{A.~M. Rappe}, \bibinfo{author}{C.~Kane},
  \bibinfo{author}{B.~A. Bernevig}.
\newblock \bibinfo{title}{Wallpaper fermions and the nonsymmorphic {Dirac}
  insulator}.
\newblock \emph{\bibinfo{journal}{Science}} \textbf{\bibinfo{volume}{361}},
  \bibinfo{pages}{246--251} (\bibinfo{year}{2018}).

\bibitem{wallpaperjacs}
\bibinfo{author}{D.-C. Ryu}, \bibinfo{author}{J.~Kim},
  \bibinfo{author}{H.~Choi}, \bibinfo{author}{B.~I. Min}.
\newblock \bibinfo{title}{Wallpaper {Dirac} fermion in a nonsymmorphic
  topological kondo insulator: {PuB$_4$}}.
\newblock \emph{\bibinfo{journal}{J. Am. Chem. Soc.}}
  \textbf{\bibinfo{volume}{142}}, \bibinfo{pages}{19278--19282}
  (\bibinfo{year}{2020}).

\bibitem{mobiustwist}
\bibinfo{author}{K.~Shiozaki}, \bibinfo{author}{M.~Sato},
  \bibinfo{author}{K.~Gomi}.
\newblock \bibinfo{title}{Z$_2$ topology in nonsymmorphic crystalline
  insulators: M{\"o}bius twist in surface states}.
\newblock \emph{\bibinfo{journal}{Phys. Rev. B}} \textbf{\bibinfo{volume}{91}},
  \bibinfo{pages}{155120} (\bibinfo{year}{2015}).

\bibitem{offcentered}
\bibinfo{author}{B.-J. Yang}, \bibinfo{author}{T.~A. Bojesen},
  \bibinfo{author}{T.~Morimoto}, \bibinfo{author}{A.~Furusaki}.
\newblock \bibinfo{title}{Topological semimetals protected by off-centered
  symmetries in nonsymmorphic crystals}.
\newblock \emph{\bibinfo{journal}{Phys. Rev. B}} \textbf{\bibinfo{volume}{95}},
  \bibinfo{pages}{075135} (\bibinfo{year}{2017}).

\bibitem{fangnonsymmor}
\bibinfo{author}{C.~Fang}, \bibinfo{author}{L.~Fu}.
\newblock \bibinfo{title}{New classes of three-dimensional topological
  crystalline insulators: Nonsymmorphic and magnetic}.
\newblock \emph{\bibinfo{journal}{Phys. Rev. B}} \textbf{\bibinfo{volume}{91}},
  \bibinfo{pages}{161105(R)} (\bibinfo{year}{2015}).

\bibitem{nonsymmorphichigher}
\bibinfo{author}{Y.~Liu}, \bibinfo{author}{Y.~Wang}, \bibinfo{author}{N.~C.
  Hu}, \bibinfo{author}{J.~Y. Lin}, \bibinfo{author}{C.~H. Lee},
  \bibinfo{author}{X.~Zhang}.
\newblock \bibinfo{title}{Topological corner modes in a brick lattice with
  nonsymmorphic symmetry}.
\newblock \emph{\bibinfo{journal}{Phys. Rev. B}}
  \textbf{\bibinfo{volume}{102}}, \bibinfo{pages}{035142}
  (\bibinfo{year}{2020}).

\bibitem{liucxnonsymmor}
\bibinfo{author}{C.-X. Liu}, \bibinfo{author}{R.-X. Zhang},
  \bibinfo{author}{B.~K. VanLeeuwen}.
\newblock \bibinfo{title}{Topological nonsymmorphic crystalline insulators}.
\newblock \emph{\bibinfo{journal}{Phys. Rev. B}} \textbf{\bibinfo{volume}{90}},
  \bibinfo{pages}{085304} (\bibinfo{year}{2014}).

\bibitem{Benalcazar61}
\bibinfo{author}{W.~A. Benalcazar}, \bibinfo{author}{B.~A. Bernevig},
  \bibinfo{author}{T.~L. Hughes}.
\newblock \bibinfo{title}{Quantized electric multipole insulators}.
\newblock \emph{\bibinfo{journal}{Science}} \textbf{\bibinfo{volume}{357}},
  \bibinfo{pages}{61--66} (\bibinfo{year}{2017}).

\bibitem{benalcazar2017electric}
\bibinfo{author}{W.~A. Benalcazar}, \bibinfo{author}{B.~A. Bernevig},
  \bibinfo{author}{T.~L. Hughes}.
\newblock \bibinfo{title}{Electric multipole moments, topological multipole
  moment pumping, and chiral hinge states in crystalline insulators}.
\newblock \emph{\bibinfo{journal}{Phys. Rev. B}} \textbf{\bibinfo{volume}{96}},
  \bibinfo{pages}{245115} (\bibinfo{year}{2017}).

\bibitem{highorderfirst}
\bibinfo{author}{F.~Schindler}, \bibinfo{author}{A.~M. Cook},
  \bibinfo{author}{M.~G. Vergniory}, \bibinfo{author}{Z.~Wang},
  \bibinfo{author}{S.~S. Parkin}, \bibinfo{author}{B.~A. Bernevig},
  \bibinfo{author}{T.~Neupert}.
\newblock \bibinfo{title}{Higher-order topological insulators}.
\newblock \emph{\bibinfo{journal}{Sci. Adv.}} \textbf{\bibinfo{volume}{4}},
  \bibinfo{pages}{eaat0346} (\bibinfo{year}{2018}).

\bibitem{ni2020demonstration}
\bibinfo{author}{X.~Ni}, \bibinfo{author}{M.~Li}, \bibinfo{author}{M.~Weiner},
  \bibinfo{author}{A.~Al{\`u}}, \bibinfo{author}{A.~B. Khanikaev}.
\newblock \bibinfo{title}{Demonstration of a quantized acoustic octupole
  topological insulator}.
\newblock \emph{\bibinfo{journal}{Nat. Commun.}} \textbf{\bibinfo{volume}{11}},
  \bibinfo{pages}{1--7} (\bibinfo{year}{2020}).

\bibitem{highordergranpj}
\bibinfo{author}{E.~Lee}, \bibinfo{author}{R.~Kim}, \bibinfo{author}{J.~Ahn},
  \bibinfo{author}{B.-J. Yang}.
\newblock \bibinfo{title}{Two-dimensional higher-order topology in monolayer
  graphdiyne}.
\newblock \emph{\bibinfo{journal}{npj Quantum Mater.}}
  \textbf{\bibinfo{volume}{5}}, \bibinfo{pages}{1--7} (\bibinfo{year}{2020}).

\bibitem{higherordereuln2as2}
\bibinfo{author}{Y.~Xu}, \bibinfo{author}{Z.~Song}, \bibinfo{author}{Z.~Wang},
  \bibinfo{author}{H.~Weng}, \bibinfo{author}{X.~Dai}.
\newblock \bibinfo{title}{Higher-order topology of the axion insulator
  {EuIn$_2$As$_2$}}.
\newblock \emph{\bibinfo{journal}{Phys. Rev. Lett.}}
  \textbf{\bibinfo{volume}{122}}, \bibinfo{pages}{256402}
  (\bibinfo{year}{2019}).

\bibitem{Garcia2018}
\bibinfo{author}{M.~Serra-Garcia}, \bibinfo{author}{V.~Peri},
  \bibinfo{author}{R.~S{\"{u}}sstrunk}, \bibinfo{author}{O.~R. Bilal},
  \bibinfo{author}{T.~Larsen}, \bibinfo{author}{L.~G. Villanueva},
  \bibinfo{author}{S.~D. Huber}.
\newblock \bibinfo{title}{{Observation of a phononic quadrupole topological
  insulator}}.
\newblock \emph{\bibinfo{journal}{Nature}} \textbf{\bibinfo{volume}{555}},
  \bibinfo{pages}{342--345} (\bibinfo{year}{2018}).

\bibitem{Imhof2018}
\bibinfo{author}{S.~Imhof}, \bibinfo{author}{C.~Berger},
  \bibinfo{author}{F.~Bayer}, \bibinfo{author}{J.~Brehm},
  \bibinfo{author}{L.~W. Molenkamp}, \bibinfo{author}{T.~Kiessling},
  \bibinfo{author}{F.~Schindler}, \bibinfo{author}{C.~H. Lee},
  \bibinfo{author}{M.~Greiter}, \bibinfo{author}{T.~Neupert},
  \bibinfo{author}{R.~Thomale}.
\newblock \bibinfo{title}{{Topolectrical-circuit realization of topological
  corner modes}}.
\newblock \emph{\bibinfo{journal}{Nat. Phys.}} \textbf{\bibinfo{volume}{14}},
  \bibinfo{pages}{925--929} (\bibinfo{year}{2018}).

\bibitem{Peterson2018}
\bibinfo{author}{C.~W. Peterson}, \bibinfo{author}{W.~A. Benalcazar},
  \bibinfo{author}{T.~L. Hughes}, \bibinfo{author}{G.~Bahl}.
\newblock \bibinfo{title}{{A quantized microwave quadrupole insulator with
  topologically protected corner states}}.
\newblock \emph{\bibinfo{journal}{Nature}} \textbf{\bibinfo{volume}{555}},
  \bibinfo{pages}{346--350} (\bibinfo{year}{2018}).

\bibitem{xue2019acoustic}
\bibinfo{author}{H.~Xue}, \bibinfo{author}{Y.~Yang}, \bibinfo{author}{F.~Gao},
  \bibinfo{author}{Y.~Chong}, \bibinfo{author}{B.~Zhang}.
\newblock \bibinfo{title}{Acoustic higher-order topological insulator on a
  kagome lattice}.
\newblock \emph{\bibinfo{journal}{Nat. Mater.}} \textbf{\bibinfo{volume}{18}},
  \bibinfo{pages}{108--112} (\bibinfo{year}{2019}).

\bibitem{ni2019observation}
\bibinfo{author}{X.~Ni}, \bibinfo{author}{M.~Weiner}, \bibinfo{author}{A.~Alu},
  \bibinfo{author}{A.~B. Khanikaev}.
\newblock \bibinfo{title}{Observation of higher-order topological acoustic
  states protected by generalized chiral symmetry}.
\newblock \emph{\bibinfo{journal}{Nat. Mater.}} \textbf{\bibinfo{volume}{18}},
  \bibinfo{pages}{113--120} (\bibinfo{year}{2019}).

\bibitem{highorderbieuo}
\bibinfo{author}{C.~Chen}, \bibinfo{author}{Z.~Song}, \bibinfo{author}{J.-Z.
  Zhao}, \bibinfo{author}{Z.~Chen}, \bibinfo{author}{Z.-M. Yu},
  \bibinfo{author}{X.-L. Sheng}, \bibinfo{author}{S.~A. Yang}.
\newblock \bibinfo{title}{Universal approach to magnetic second-order
  topological insulator}.
\newblock \emph{\bibinfo{journal}{Phys. Rev. Lett.}}
  \textbf{\bibinfo{volume}{125}}, \bibinfo{pages}{056402}
  (\bibinfo{year}{2020}).

\bibitem{Yue2019}
\bibinfo{author}{C.~Yue}, \bibinfo{author}{Y.~Xu}, \bibinfo{author}{Z.~Song},
  \bibinfo{author}{H.~Weng}, \bibinfo{author}{Y.~M. Lu},
  \bibinfo{author}{C.~Fang}, \bibinfo{author}{X.~Dai}.
\newblock \bibinfo{title}{{Symmetry-enforced chiral hinge states and surface
  quantum anomalous Hall effect in the magnetic axion insulator
  Bi$_{2-x}$Sm$_x$Se$_3$}}.
\newblock \emph{\bibinfo{journal}{Nat. Phys.}} \textbf{\bibinfo{volume}{15}},
  \bibinfo{pages}{577--581} (\bibinfo{year}{2019}).

\bibitem{mnbiteprl1}
\bibinfo{author}{D.~Zhang}, \bibinfo{author}{M.~Shi}, \bibinfo{author}{T.~Zhu},
  \bibinfo{author}{D.~Xing}, \bibinfo{author}{H.~Zhang},
  \bibinfo{author}{J.~Wang}.
\newblock \bibinfo{title}{Topological axion states in the magnetic insulator
  {MnBi$_2$Te$_4$} with the quantized magnetoelectric effect}.
\newblock \emph{\bibinfo{journal}{Phys. Rev. Lett.}}
  \textbf{\bibinfo{volume}{122}}, \bibinfo{pages}{206401}
  (\bibinfo{year}{2019}).

\bibitem{mnbitenm1}
\bibinfo{author}{C.~Liu}, \bibinfo{author}{Y.~Wang}, \bibinfo{author}{H.~Li},
  \bibinfo{author}{Y.~Wu}, \bibinfo{author}{Y.~Li}, \bibinfo{author}{J.~Li},
  \bibinfo{author}{K.~He}, \bibinfo{author}{Y.~Xu}, \bibinfo{author}{J.~Zhang},
  \bibinfo{author}{Y.~Wang}.
\newblock \bibinfo{title}{Robust axion insulator and chern insulator phases in
  a two-dimensional antiferromagnetic topological insulator}.
\newblock \emph{\bibinfo{journal}{Nat. Mater.}} \textbf{\bibinfo{volume}{19}},
  \bibinfo{pages}{522--527} (\bibinfo{year}{2020}).

\bibitem{nature4}
\bibinfo{author}{Y.~Xu}, \bibinfo{author}{L.~Elcoro}, \bibinfo{author}{Z.~D.
  Song}, \bibinfo{author}{B.~J. Wieder}, \bibinfo{author}{M.~G. Vergniory},
  \bibinfo{author}{N.~Regnault}, \bibinfo{author}{Y.~Chen},
  \bibinfo{author}{C.~Felser}, \bibinfo{author}{B.~A. Bernevig}.
\newblock \bibinfo{title}{{High-throughput calculations of magnetic topological
  materials}}.
\newblock \emph{\bibinfo{journal}{Nature}} \textbf{\bibinfo{volume}{586}},
  \bibinfo{pages}{702--707} (\bibinfo{year}{2020}).

\bibitem{Park216803}
\bibinfo{author}{M.~J. Park}, \bibinfo{author}{Y.~Kim}, \bibinfo{author}{G.~Y.
  Cho}, \bibinfo{author}{S.~Lee}.
\newblock \bibinfo{title}{Higher-order topological insulator in twisted bilayer
  graphene}.
\newblock \emph{\bibinfo{journal}{Phys. Rev. Lett.}}
  \textbf{\bibinfo{volume}{123}}, \bibinfo{pages}{216803}
  (\bibinfo{year}{2019}).

\bibitem{frankbi}
\bibinfo{author}{F.~Schindler}, \bibinfo{author}{Z.~Wang},
  \bibinfo{author}{M.~G. Vergniory}, \bibinfo{author}{A.~M. Cook},
  \bibinfo{author}{A.~Murani}, \bibinfo{author}{S.~Sengupta},
  \bibinfo{author}{A.~Y. Kasumov}, \bibinfo{author}{R.~Deblock},
  \bibinfo{author}{S.~Jeon}, \bibinfo{author}{I.~Drozdov},
  \bibinfo{author}{H.~Bouchiat}, \bibinfo{author}{S.~Gu{\'{e}}ron},
  \bibinfo{author}{A.~Yazdani}, \bibinfo{author}{B.~A. Bernevig},
  \bibinfo{author}{T.~Neupert}.
\newblock \bibinfo{title}{{Higher-order topology in bismuth}}.
\newblock \emph{\bibinfo{journal}{Nat. Phys.}} \textbf{\bibinfo{volume}{14}},
  \bibinfo{pages}{918--924} (\bibinfo{year}{2018}).

\bibitem{kempkes2019robust}
\bibinfo{author}{S.~Kempkes}, \bibinfo{author}{M.~Slot},
  \bibinfo{author}{J.~van Den~Broeke}, \bibinfo{author}{P.~Capiod},
  \bibinfo{author}{W.~Benalcazar}, \bibinfo{author}{D.~Vanmaekelbergh},
  \bibinfo{author}{D.~Bercioux}, \bibinfo{author}{I.~Swart},
  \bibinfo{author}{C.~M. Smith}.
\newblock \bibinfo{title}{Robust zero-energy modes in an electronic
  higher-order topological insulator}.
\newblock \emph{\bibinfo{journal}{Nat. Mater.}} \textbf{\bibinfo{volume}{18}},
  \bibinfo{pages}{1292--1297} (\bibinfo{year}{2019}).

\bibitem{noguchi2021evidence}
\bibinfo{author}{R.~Noguchi}, \bibinfo{author}{M.~Kobayashi},
  \bibinfo{author}{Z.~Jiang}, \bibinfo{author}{K.~Kuroda},
  \bibinfo{author}{T.~Takahashi}, \bibinfo{author}{Z.~Xu},
  \bibinfo{author}{D.~Lee}, \bibinfo{author}{M.~Hirayama},
  \bibinfo{author}{M.~Ochi}, \bibinfo{author}{T.~Shirasawa}, et~al.
\newblock \bibinfo{title}{Evidence for a higher-order topological insulator in
  a three-dimensional material built from van der waals stacking of
  bismuth-halide chains}.
\newblock \emph{\bibinfo{journal}{Nat. Mater.}} \textbf{\bibinfo{volume}{20}},
  \bibinfo{pages}{473--479} (\bibinfo{year}{2021}).

\bibitem{highordergraprl}
\bibinfo{author}{X.-L. Sheng}, \bibinfo{author}{C.~Chen},
  \bibinfo{author}{H.~Liu}, \bibinfo{author}{Z.~Chen}, \bibinfo{author}{Z.-M.
  Yu}, \bibinfo{author}{Y.~Zhao}, \bibinfo{author}{S.~A. Yang}.
\newblock \bibinfo{title}{Two-dimensional second-order topological insulator in
  graphdiyne}.
\newblock \emph{\bibinfo{journal}{Phys. Rev. Lett.}}
  \textbf{\bibinfo{volume}{123}}, \bibinfo{pages}{256402}
  (\bibinfo{year}{2019}).

\bibitem{highordersplit}
\bibinfo{author}{B.~Liu}, \bibinfo{author}{G.~Zhao}, \bibinfo{author}{Z.~Liu},
  \bibinfo{author}{Z.~Wang}.
\newblock \bibinfo{title}{Two-dimensional quadrupole topological insulator in
  $\gamma$-graphyne}.
\newblock \emph{\bibinfo{journal}{Nano Lett.}} \textbf{\bibinfo{volume}{19}},
  \bibinfo{pages}{6492--6497} (\bibinfo{year}{2019}).

\bibitem{highorderinvariants}
\bibinfo{author}{W.~A. Benalcazar}, \bibinfo{author}{T.~Li},
  \bibinfo{author}{T.~L. Hughes}.
\newblock \bibinfo{title}{Quantization of fractional corner charge in
  {C$_n$-symmetric} higher-order topological crystalline insulators}.
\newblock \emph{\bibinfo{journal}{Phys. Rev. B}} \textbf{\bibinfo{volume}{99}},
  \bibinfo{pages}{245151} (\bibinfo{year}{2019}).

\bibitem{highordersonginvariants}
\bibinfo{author}{Z.~Song}, \bibinfo{author}{Z.~Fang},
  \bibinfo{author}{C.~Fang}.
\newblock \bibinfo{title}{(d - 2)-dimensional edge states of rotation symmetry
  protected topological states}.
\newblock \emph{\bibinfo{journal}{Phys. Rev. Lett.}}
  \textbf{\bibinfo{volume}{119}}, \bibinfo{pages}{246402}
  (\bibinfo{year}{2017}).

\bibitem{Peterson1114}
\bibinfo{author}{C.~W. Peterson}, \bibinfo{author}{T.~Li},
  \bibinfo{author}{W.~A. Benalcazar}, \bibinfo{author}{T.~L. Hughes},
  \bibinfo{author}{G.~Bahl}.
\newblock \bibinfo{title}{A fractional corner anomaly reveals higher-order
  topology}.
\newblock \emph{\bibinfo{journal}{Science}} \textbf{\bibinfo{volume}{368}},
  \bibinfo{pages}{1114--1118} (\bibinfo{year}{2020}).

\bibitem{Xue2019PRL}
\bibinfo{author}{H.~Xue}, \bibinfo{author}{Y.~Yang}, \bibinfo{author}{G.~Liu},
  \bibinfo{author}{F.~Gao}, \bibinfo{author}{Y.~Chong},
  \bibinfo{author}{B.~Zhang}.
\newblock \bibinfo{title}{Realization of an acoustic third-order topological
  insulator}.
\newblock \emph{\bibinfo{journal}{Phys. Rev. Lett.}}
  \textbf{\bibinfo{volume}{122}}, \bibinfo{pages}{244301}
  (\bibinfo{year}{2019}).

\bibitem{liu2020octupole}
\bibinfo{author}{S.~Liu}, \bibinfo{author}{S.~Ma}, \bibinfo{author}{Q.~Zhang},
  \bibinfo{author}{L.~Zhang}, \bibinfo{author}{C.~Yang},
  \bibinfo{author}{O.~You}, \bibinfo{author}{W.~Gao},
  \bibinfo{author}{Y.~Xiang}, \bibinfo{author}{T.~J. Cui},
  \bibinfo{author}{S.~Zhang}.
\newblock \bibinfo{title}{Octupole corner state in a three-dimensional
  topological circuit}.
\newblock \emph{\bibinfo{journal}{Light Sci. Appl.}}
  \textbf{\bibinfo{volume}{9}}, \bibinfo{pages}{1--9} (\bibinfo{year}{2020}).

\bibitem{tl4te3x4}
\bibinfo{author}{K.~Arpino}, \bibinfo{author}{D.~Wallace},
  \bibinfo{author}{Y.~Nie}, \bibinfo{author}{T.~Birol},
  \bibinfo{author}{P.~King}, \bibinfo{author}{S.~Chatterjee},
  \bibinfo{author}{M.~Uchida}, \bibinfo{author}{S.~Koohpayeh},
  \bibinfo{author}{J.-J. Wen}, \bibinfo{author}{K.~Page}, et~al.
\newblock \bibinfo{title}{Evidence for topologically protected surface states
  and a superconducting phase in {[Tl$_4$](Tl$_{1-x}$Sn$_x$)Te$_3$} using
  photoemission, specific heat, and magnetization measurements, and density
  functional theory}.
\newblock \emph{\bibinfo{journal}{Phys. Rev. Lett.}}
  \textbf{\bibinfo{volume}{112}}, \bibinfo{pages}{017002}
  (\bibinfo{year}{2014}).

\bibitem{nature1}
\bibinfo{author}{M.~Vergniory}, \bibinfo{author}{L.~Elcoro},
  \bibinfo{author}{C.~Felser}, \bibinfo{author}{N.~Regnault},
  \bibinfo{author}{B.~A. Bernevig}, \bibinfo{author}{Z.~Wang}.
\newblock \bibinfo{title}{A complete catalogue of high-quality topological
  materials}.
\newblock \emph{\bibinfo{journal}{Nature}} \textbf{\bibinfo{volume}{566}},
  \bibinfo{pages}{480--485} (\bibinfo{year}{2019}).

\bibitem{nature2}
\bibinfo{author}{T.~Zhang}, \bibinfo{author}{Y.~Jiang},
  \bibinfo{author}{Z.~Song}, \bibinfo{author}{H.~Huang},
  \bibinfo{author}{Y.~He}, \bibinfo{author}{Z.~Fang},
  \bibinfo{author}{H.~Weng}, \bibinfo{author}{C.~Fang}.
\newblock \bibinfo{title}{Catalogue of topological electronic materials}.
\newblock \emph{\bibinfo{journal}{Nature}} \textbf{\bibinfo{volume}{566}},
  \bibinfo{pages}{475--479} (\bibinfo{year}{2019}).

\bibitem{lcxtnci}
\bibinfo{author}{C.-X. Liu}, \bibinfo{author}{R.-X. Zhang},
  \bibinfo{author}{B.~K. VanLeeuwen}.
\newblock \bibinfo{title}{Topological nonsymmorphic crystalline insulators}.
\newblock \emph{\bibinfo{journal}{Phys. Rev. B}} \textbf{\bibinfo{volume}{90}},
  \bibinfo{pages}{085304} (\bibinfo{year}{2014}).

\bibitem{tl4te3x1}
\bibinfo{author}{A.~Kosuga}, \bibinfo{author}{K.~Kurosaki},
  \bibinfo{author}{H.~Muta}, \bibinfo{author}{S.~Yamanaka}.
\newblock \bibinfo{title}{{Thermoelectric properties of Tl-X-Te (X= Pb, Sn, Ge)
  systems}}.
\newblock \emph{\bibinfo{journal}{Mater. Res. Soc. Symp. Proc.}}
  \textbf{\bibinfo{volume}{886}} (\bibinfo{year}{2005}).

\bibitem{tl4te3x3}
\bibinfo{author}{I.~Barchij}, \bibinfo{author}{M.~Sabov},
  \bibinfo{author}{A.~El-Naggar}, \bibinfo{author}{N.~AlZayed},
  \bibinfo{author}{A.~Albassam}, \bibinfo{author}{A.~Fedorchuk},
  \bibinfo{author}{I.~Kityk}.
\newblock \bibinfo{title}{{Tl$_4$SnS$_3$, Tl$_4$SnSe$_3$ and Tl$_4$SnTe$_3$
  crystals as novel IR induced optoelectronic materials}}.
\newblock \emph{\bibinfo{journal}{J. Mater. Sci. Mater. Electron.}}
  \textbf{\bibinfo{volume}{27}}, \bibinfo{pages}{3901--3905}
  (\bibinfo{year}{2016}).

\bibitem{tl4te3x2}
\bibinfo{author}{M.~Filep}, \bibinfo{author}{M.~Y. Sabov},
  \bibinfo{author}{I.~Barchiy}, \bibinfo{author}{K.~Plucinski},
  \bibinfo{author}{A.~Solomon}.
\newblock \bibinfo{title}{{Interactions in the ternary reciprocal system
  Tl$_2$S + SnTe $\leftrightarrow$Tl$_2$Te+ SnS}}.
\newblock \emph{\bibinfo{journal}{Chemistry of metals and alloys}}
  \bibinfo{pages}{125--129} (\bibinfo{year}{2013}).

\bibitem{wyckoff}
\bibinfo{author}{M.~I. Aroyo}, \bibinfo{author}{J.~M. Perez-Mato},
  \bibinfo{author}{C.~Capillas}, \bibinfo{author}{E.~Kroumova},
  \bibinfo{author}{S.~Ivantchev}, \bibinfo{author}{G.~Madariaga},
  \bibinfo{author}{A.~Kirov}, \bibinfo{author}{H.~Wondratschek}.
\newblock \bibinfo{title}{Bilbao crystallographic server: I. databases and
  crystallographic computing programs}.
\newblock \emph{\bibinfo{journal}{Z. Kristallogr. Cryst. Mater.}}
  \textbf{\bibinfo{volume}{221}}, \bibinfo{pages}{15--27}
  (\bibinfo{year}{2006}).

\bibitem{z2invariant}
\bibinfo{author}{L.~Fu}, \bibinfo{author}{C.~L. Kane}.
\newblock \bibinfo{title}{Topological insulators with inversion symmetry}.
\newblock \emph{\bibinfo{journal}{Phys. Rev. B}} \textbf{\bibinfo{volume}{76}},
  \bibinfo{pages}{045302} (\bibinfo{year}{2007}).

\bibitem{z2bump}
\bibinfo{author}{L.~Fu}, \bibinfo{author}{C.~L. Kane}.
\newblock \bibinfo{title}{Time reversal polarization and a {Z$_2$} adiabatic
  spin pump}.
\newblock \emph{\bibinfo{journal}{Phys. Rev. B}} \textbf{\bibinfo{volume}{74}},
  \bibinfo{pages}{195312} (\bibinfo{year}{2006}).

\bibitem{pbevasp}
\bibinfo{author}{J.~P. Perdew}, \bibinfo{author}{K.~Burke},
  \bibinfo{author}{M.~Ernzerhof}.
\newblock \bibinfo{title}{Generalized gradient approximation made simple}.
\newblock \emph{\bibinfo{journal}{Phys. Rev. Lett.}}
  \textbf{\bibinfo{volume}{77}}, \bibinfo{pages}{3865} (\bibinfo{year}{1996}).

\bibitem{vasp}
\bibinfo{author}{G.~Kresse}, \bibinfo{author}{J.~Furthm{\"{u}}ller}.
\newblock \bibinfo{title}{{Efficient iterative schemes for ab initio
  total-energy calculations using a plane-wave basis set}}.
\newblock \emph{\bibinfo{journal}{Phys. Rev. B}} \textbf{\bibinfo{volume}{54}},
  \bibinfo{pages}{11169--11186} (\bibinfo{year}{1996}).

\bibitem{fleur}

\newblock \bibinfo{howpublished}{See \url{http://www.flapw.de}}.

\bibitem{wannier90}
\bibinfo{author}{G.~Pizzi}, \bibinfo{author}{V.~Vitale},
  \bibinfo{author}{R.~Arita}, \bibinfo{author}{S.~Bl{\"u}gel},
  \bibinfo{author}{F.~Freimuth}, \bibinfo{author}{G.~G{\'e}ranton},
  \bibinfo{author}{M.~Gibertini}, \bibinfo{author}{D.~Gresch},
  \bibinfo{author}{C.~Johnson}, \bibinfo{author}{T.~Koretsune}, et~al.
\newblock \bibinfo{title}{Wannier90 as a community code: {New} features and
  applications}.
\newblock \emph{\bibinfo{journal}{J. Phys. Condens. Matter}}
  \textbf{\bibinfo{volume}{32}}, \bibinfo{pages}{165902}
  (\bibinfo{year}{2020}).

\bibitem{fluer1}
\bibinfo{author}{A.~A. Mostofi}, \bibinfo{author}{J.~R. Yates},
  \bibinfo{author}{Y.-S. Lee}, \bibinfo{author}{I.~Souza},
  \bibinfo{author}{D.~Vanderbilt}, \bibinfo{author}{N.~Marzari}.
\newblock \bibinfo{title}{Wannier90: {A} tool for obtaining maximally-localised
  wannier functions}.
\newblock \emph{\bibinfo{journal}{Comput. Phys. Commun.}}
  \textbf{\bibinfo{volume}{178}}, \bibinfo{pages}{685 -- 699}
  (\bibinfo{year}{2008}).

\bibitem{fluer2}
\bibinfo{author}{F.~Freimuth}, \bibinfo{author}{Y.~Mokrousov},
  \bibinfo{author}{D.~Wortmann}, \bibinfo{author}{S.~Heinze},
  \bibinfo{author}{S.~Bl{\"{u}}gel}.
\newblock \bibinfo{title}{{Maximally localized Wannier functions within the
  {F}{L}{A}{P}{W} formalism}}.
\newblock \emph{\bibinfo{journal}{Phys. Rev. B}} \textbf{\bibinfo{volume}{78}},
  \bibinfo{pages}{035120} (\bibinfo{year}{2008}).

\bibitem{wanniertools}
\bibinfo{author}{Q.~Wu}, \bibinfo{author}{S.~Zhang}, \bibinfo{author}{H.-F.
  Song}, \bibinfo{author}{M.~Troyer}, \bibinfo{author}{A.~A. Soluyanov}.
\newblock \bibinfo{title}{Wanniertools: {An} open-source software package for
  novel topological materials}.
\newblock \emph{\bibinfo{journal}{Comput. Phys. Commun.}}
  \textbf{\bibinfo{volume}{224}}, \bibinfo{pages}{405--416}
  (\bibinfo{year}{2018}).

\end{thebibliography}

\vspace{0.8cm}
\noindent{\bf \textcolor{black}{Acknowledgements}}
This work was supported by the National Natural Science Foundation of China (Grants No. 11904205 and No. 12074217), the Shandong Provincial Natural Science Foundation of China (Grants No. ZR2019QA019 and No. ZR2019MEM013), the Shandong Provincial Key Research and Development Program (Major Scientific and Technological Innovation Project) (Grant No. 2019JZZY010302), the Taishan Scholar Program of Shandong Province, and the Qilu Young Scholar Program of Shandong University. {\bf \textcolor{black}{Author contributions}}:C.N. and Y.D. conceived the project. N.M. performed the first-principles calculations and model analysis with the help of H.W. and C.N.. N.M. and C.N. wrote the manuscript with contributions from B.H. and Y.D.. All authors discussed the results and commented on the manuscript. {\bf \textcolor{black}{Competing interests}}: The authors declare that they have no competing interests. {\bf \textcolor{black}{Data and materials availability}}: All data needed to evaluate the conclusions in the paper are present in the paper and/or the Supplementary Materials. Additional data related to this paper may be requested from the authors.

\end{document}